\documentclass[12pt]{article}
\usepackage{epsfig,amsmath}
  {\end{list}}%

\def\cA{{\cal A}}
\def\cJ{{\cal J}}
\def\cO{{\cal O}}
\def\cQ{{\cal Q}}
\def\cX{{\cal X}}

\def\A{{\rm A}}
\def\B{{\rm B}}
\def\C{{\rm C}}
\def\E{{\rm E}}
\def\F{{\rm F}}
\def\W{{\rm W}}
\def\Z{{\rm Z}}

\def\Aslash{{\A\mkern-12mu/}}
\def\Bslash{{\B\mkern-12mu/}}
\def\Dirac{{D\mkern-12mu/}}
\def\prslash{{\partial\mkern-9mu/}}

\def\prslash{{\partial\mkern-9mu/}}    

\def\idpn{\int\! \frac{d^{2n}\!p}{(2\pi)^{2n}} \,\,}

\def\idqn{\int\! \frac{d^{2n}\!q}{(2\pi)^{2n}} \,\,}

\def\idxn{\int\! d^{2n}\!x \,}

\def\ot{\otimes}
\def\ds{\displaystyle}
\def\RR{{\rm I\!\!\, R}}
\def\MM{{\rm I\!\!\, M}}
\def\unit{{\rm I\!\!\, I}}
\def\bracl{ [\!\!\, [}
\def\bracr{ ]\!\!\, ]}

\begin{document}
\begin{titlepage}
\rightline{UCM-FT/00-65-2001}

\vskip 1.5 true cm
\begin{center}
{\Large \bf The covariant form of the gauge anomaly\\[9pt]
on noncommutative $\RR^{2 n}$}\\ 
\vskip 1.2 true cm 
{\rm C.P. Mart\'{\i}n}\footnote{E-mail: carmelo@elbereth.fis.ucm.es}
\vskip 0.3 true cm
{\it Departamento de F\'{\i}sica Te\'orica I}\\
{\it Facultad de Ciencias F\'{\i}sicas}\\ 
{\it Universidad Complutense de Madrid}\\
{\it 28040 Madrid, Spain}\\
\vskip 1.2 true cm

{\leftskip=45pt \rightskip=45pt 
\noindent
The covariant form of the non-Abelian gauge anomaly on noncommutative 
$\RR^{2 n}$ is computed for $U(N)$ groups. Its origin and  properties  
are analyzed.  
Its connection with the consistent form of the gauge anomaly is established. 
We show along the way that bi-fundamental $U(N)\times U(M)$ chiral matter 
carries no mixed anomalies, and interpret this result as a consequence of 
the half-dipole structure which  characterizes the charged non-commutative 
degrees of freedom. \par }
\end{center}

\vfil

\end{titlepage}
\setcounter{page}{2}


\section{Introduction}

Field theories of fermions with chiral couplings to gauge fields on  
commutative manifolds play a prominent role --at least up to a few Tev-- in 
the description of Nature.
Field theories over noncommutative 
space-time~\cite{Douglas:2001ba, Szabo:2001kg} may 
turn out to be phenomenologically relevant at the Tev scale and 
above~\cite{Riad:2000vy, Arfaei:2000kh, Hewett:2001zp, Mathews:2001we, Baek:2001ty, Grosse:2001ei, Grosse:2001xz, 
Chaichian:2001py, Wang:2001ig}.
It is therefore a must to understand the properties of quantum 
field theories of fermions chirally coupled to gauge fields on
noncommutative manifolds. See refs.~\cite{Connes:1994, Madore:1999, Landi:1997sh, Gracia-Bondia:2001tr} for the mathematics of noncommutative manifolds.

The chief feature of quantum field theories of chiral fermions interacting 
with gauge fields is that they are liable to carry gauge anomalies --other 
types of anomalies such the conformal anomaly~\cite{Nakajima:2001uh} will not 
be discussed here.
It is a well 
established fact~\cite{Alvarez-Gaume:1984cs, Fujikawa:1984bg, Bardeen:1984pm,  
Alvarez-Gaume:1985dr, Fujikawa:1985pt, Banerjee:1986bu, Banerjee:1999up}
 that if space-time is  
commutative, a gauge anomaly comes in either of two guises, namely,
its consistent form or its covariant form. The consistent form of the anomaly
satisfies the Wess-Zumino consistency conditions~\cite{Wess:1971yu},  
the covariant form does not.
One can retrieve  either form of the anomaly from the other by adding 
to the corresponding current a polynomial of the gauge fields and its 
derivatives. For noncommutative $\RR^4$ the consistent form of the gauge 
anomaly has been obtained in a number of papers~\cite{Gracia-Bondia:2000pz, Bonora:2000he, Grisaru:2001sk} --see also 
ref.~\cite{Langmann:1995ub, Perrot:1999vc, Harvey:2001pd} for general 
analysis of chiral anomalies on noncommutative spaces and 
ref.~\cite{Ardalan:2001cy, Ardalan:2000qk} for explicit computations. 
As for its covariant form, there is as yet no thorough discussion of the
gauge anomaly for noncommutative space-time --although some results have 
been issued in ref.~\cite{Martin:2001wc}. The purpose of this paper is to
remedy this situation. First, by using path integral techniques,  we shall   
compute explicitly the form of the gauge anomaly on noncommutative 
$\RR^{2 n}$ for $U(N)$ groups. In so doing, we shall see that the covariant 
form of the gauge anomaly is associated with a given 
definition of the path integral. This definition being a $*\!$-deformation of
the ordinary one as given in ref.~\cite{Fujikawa:1984bg, Banerjee:1986bu, 
 Banerjee:1999up}. Then, we
shall show that the covariant form of the gauge anomaly can be turned 
into the  consistent form of it, by adding to the covariant 
current a $*\!$-polynomial of the gauge field and the field strength; this 
$*\!$-polynomial been a $*\!$-deformation of the polynomial for the commutative 
$\RR^{2 n}$ case. Finally, we shall analyze the transformation properties, 
under gauge transformations of the gauge field, of both
the  both the consistent and covariant currents. We shall thus show that 
in the presence of the gauge anomaly the consistent current --that which 
can be obtained by functional differentiation of the effective action--  cannot 
transform covariantly; whereas, the covariant current does transform 
covariantly and, hence,  it cannot be  the functional derivative  of 
the effective action with respect to the gauge field. 
For  commutative $\RR^{2 n}$, these properties of the currents were 
established in refs.~\cite{Bardeen:1984pm, Alvarez-Gaume:1985dr, Banerjee:1986bu, Banerjee:1999up}.

\section{The covariant form of the gauge anomaly: fundamental matter}

Let $\psi^j_{\rm R}$ denote a right handed fermion, 
$\psi^j_{\rm R}\!=\!{\rm P}_{+}\psi^{j}$, ${\rm P}_{+}=(1+\gamma_{2n+1})/2$,  
carrying the fundamental representation of the group $U(N)$. 
The matrix $\gamma_{2n+1}$ is given by $\gamma_{2n+1}=(-i)^{n}\prod_{\mu=1}^{2n}\,\gamma^{\mu}$, where the gamma matrices $\gamma^{\mu}$, $\mu=1,\cdots,2n$, are Hermitian matrices 
which satisfy 
$\{\gamma^{\mu},\gamma^{\nu}\}=2\delta^{\mu\nu}$. 
The physics of $\psi^j_{\rm R}$ interacting 
with a background $U(N)$  gauge field on noncommutative $\RR^{2n}$ 
is ruled by the classical action 
\begin{equation}
{\rm S}\;=\;\idxn\;\bar{\psi}_i\, i\hat{D}(\A)^i_{\;\;j}\,\psi^j.
\label{chiralclassact}
\end{equation}
The operator $i\hat{D}(\A)$, which acts on the Dirac spinor $\psi^j$ as follows 
$i\hat{D}(\A)^{i}_{\;\;j}\,\psi^j =i(\prslash\psi^{i} +
\A^{i}_{\mu\,j}\star\gamma^{\mu} {\rm P}_{+}\psi^{j})$,
is not an Hermitian operator, but it is an elliptic operator.
The Dirac spinor $\psi^j$  carries the fundamental 
representation of $U(N)$ and the complex matrix $\A^{i}_{\mu\,j}$, with 
$(\A^{i}_{\mu\,j})^*=-\A^{j}_{\mu\,i}$, is the $U(N)$ gauge field. 
The indices $i,j$ run from $1$ to $N$. The previous action is invariant under
the following chiral gauge transformations:
\begin{equation}
\begin{array}{c}
{\big(\delta_{\omega} \A_{\mu}\big)^i_{\;j}= 
-\partial_{\mu}\omega^i_{\;j}\,-\,\A^i_{\mu\,k}\star\omega^k_{\;j}\,+
\,\omega^i_{\;k}\star\A^k_{\mu\,j},}\\[9pt] 
{(\delta_{\omega}\psi)^i=\,\omega^i_{\;j}\star {\rm P}_{+}\,\psi^j,\quad
(\delta_{\omega} \bar{\psi})_{k}=
-\bar{\psi}_k\star\omega^{k}_{\;i}{\rm P}_{-},}\\[9pt]
\end{array}
\label{chiraltrans}
\end{equation}
where ${\rm P}_{-}\!=\!(1-\gamma_{2n+1})/2$. The complex functions  
$\omega^i_{\;j}=-\omega^{*\,j}_{\quad i}$, $i,j=1,\cdots, N$, 
are the infinitesimal gauge  transformation parameters. The symbol $\star$ 
denotes  the Moyal product of functions on $\RR^{2n}$.
The Moyal product is given by
$(f\star g)(x)= 
\idpn\idqn\;e^{i(p+q)_{\mu}x^{\mu}} \;
e^{-\frac{i}{2}\theta^{\mu\nu}p_{\mu}q_{\nu}}\;\hat{f}(p)\hat{g}(q)$.
Here, $\theta^{\mu\nu}$ is an anti-symmetric real matrix  either of  
magnetic type or light-like type. It is for these choices of 
matrix $\theta$ that a unitary theory exists at the quantum 
level~\cite{Gomis:2000zz, Aharony:2000gz, Alvarez-Gaume:2001ka, Bassetto:2001vf}.

To define the partition function, 
\begin{equation}
\Z[\A] \;\equiv\;\int  d\bar{\psi}d\,\psi\quad e^{-{\rm S}[\A]},
\label{partitionfunct}
\end{equation}
of the quantum theory with classical action given in 
eq.~(\ref{chiralclassact}), we shall follow Fujikawa~\cite{Fujikawa:1984bg} and use the
the set of eigenvalues and the set of eigenfunctions of the Hermitian 
operators  
$\Big(i\hat{D}(\A)\Big)^{\dagger}i\hat{D}(\A)$ and 
$i\hat{D}(\A) \Big(i\hat{D}(\A)\Big)^{\dagger}$. These sets are defined by
the following equations:
\begin{equation}
\begin{array}{l}
{\ds \Big(i\hat{D}(\A)\Big)^{\dagger}i\hat{D}(\A)\varphi_m\,=\,\lambda^2_m\,
\varphi_m,\quad i\hat{D}(\A) \Big(i\hat{D}(\A)\Big)^{\dagger}\phi_m\,=\,
\lambda^2_m\,\phi_m,}\\[9pt]
{\ds \phi_m\,=\,\frac{1}{\lambda_m}\,i\hat{D}(\A)\varphi_m,\;\;{\rm  if}\;
\lambda_m\neq 0,\;\;\; {\rm and}\;\; i\hat{D}(\A)\varphi_m\,=\,0,\;\;{\rm if}\;
\lambda_m =0},\\[9pt]
{\ds \varphi_m\,=\,\frac{1}{\lambda_m}\,\Big(i\hat{D}(\A)\Big)^{\dagger}\phi_m,\;\;}{\rm if}\;
{\ds\lambda_m\neq 0,}\;\;\; {\rm and}\;\;{\ds \;\Big(i\hat{D}(\A)\Big)^{\dagger}\phi_m\,=\,0,\;\;}{\rm if}\;
{\ds\lambda_m =0},\\[9pt]
{\ds \idxn\; \varphi_m^{\dagger}(x)\,\varphi_{m'}(x)\;=\;\delta_{mm'},\quad
 \idxn\; \phi_m^{\dagger}(x)\,\phi_{m'}(x)\;=\;\delta_{mm'}.}\\[9pt]
\end{array}
\label{definitions}
\end{equation}
We take without loss of generality $\lambda_{m}\geq 0$. We next define the 
fermionic measure as follows
\begin{equation}
d\bar{\psi}d\,\psi\quad=\prod_m\,d\bar{b}_m da_{m}.
\label{measure}
\end{equation}
Here,    
$a_m$ and $\bar{b}_m$ are Grassmann variables defined by the expansions
$\psi=\sum_{m}a_m\varphi_m$ and $\bar{\psi}=\sum_{m}\bar{b}_m\phi_m^{\dagger}$.
Notice that this definition of $d\bar{\psi}d\,\psi$ is the obvious 
generalization to the noncommutative framework of the definition in 
refs.~\cite{Banerjee:1986bu, Banerjee:1999up}. Then, 
\begin{displaymath}
\int  d\bar{\psi}d\,\psi\quad e^{-{\rm S}[\A,\psi,\bar{\psi}]}\equiv 
\int \prod_m\,d\bar{b}_m\,da_m\;e^{-\sum_m\;\lambda_m\,\bar{b}_m\,a_m};
\end{displaymath}
which after Grassmann integration leads to 
$\Z[\A]\equiv \prod_m\,\lambda_m$. Notice that we have taken into account 
eq.~(\ref{partitionfunct}). Hence, we have formally defined the 
partition function of the theory, $\Z[\A]$, as the determinant of the 
square root of the operator  
$\Big(i\hat{D}(\A)\Big)^{\dagger}i\hat{D}(\A)$. 

Now, it is not difficult to show that if $g=1+\omega$ is an infinitesimal 
gauge transformation, one has $(1+\omega{\rm P}_{+})\Big(i\hat{D}(\A^g)\Big)^{\dagger}i\hat{D}(\A^g)
 (1-\omega{\rm P}_{+})=\Big(i\hat{D}(\A)\Big)^{\dagger}i\hat{D}(\A)+
0(\omega^2)$. Hence, $\lambda_m(\A^g)=\lambda_m(\A)+O(\omega^2)$, so that
$\Z[\A]$, as defined above, is formally gauge invariant under infinitesimal 
gauge transformations. We can make this statement rigorous by using Pauli-Villars regularization or zeta function regularization. It is thus clear that 
for our definition of partition function the chiral gauge anomaly cannot be 
interpreted as the lack of invariance of $\W[\A]$ ($\W[\A]=-\ln\,\Z[\A]$)
under infinitesimal gauge transformations. An interpretation which holds 
true for the consistent form of the  
anomaly~\cite{Alvarez-Gaume:1984cs, Bardeen:1984pm}. Let us show 
next that, as in the ordinary case~\cite{Fujikawa:1984bg}, the covariant 
form of the anomaly comes
from the lack of invariance under infinitesimal chiral gauge transformations
of the fermionic measure defined above --see 
eq.~(\ref{measure}). 
Let $\psi'=\psi+\delta_{\omega}\psi$ and  
$\bar{\psi}'=\bar{\psi}+\delta_{\omega}\bar{\psi}$, where $\delta_{\omega}$ is
given in eq.~(\ref{chiraltrans}). Let $\{a'_m\}_m$ and $\{\bar{b}'_m\}_m$ be
given by the expansions  
$\psi'=\sum_ma'_m\varphi_m$ and 
$\bar{\psi}'=\sum_m\bar{b}'_m\phi_m^{\dagger}$. Then, the identity
\begin{displaymath}
\int  d\bar{\psi}d\,\psi\quad e^{-{\rm S}[\A,\psi,\bar{\psi}]}\equiv 
\int  d\bar{\psi}'d\,\psi'\quad e^{-{\rm S}[\A,\psi',\bar{\psi}']},
\end{displaymath}
leads to 
\begin{equation}
\idxn\,\omega^i_{\;j}(x)\, 
\big(D_{\mu}[\A]\,\cJ_\mu^{(cov)}\big)^{j}_{\;\,i}(x)
\,=\,-\,\delta\,J\,\equiv\,\cA[\omega,\A]^{(cov)}.
\label{formalanom}
\end{equation}
$\delta\,J$, which is defined by the equation
$\prod_m\,d\bar{b}'_m\,da'_m-
\prod_m\,d\bar{b}_m\,da_m=\delta\,J+O(\omega^2)$, is equal to
\begin{displaymath}
\idxn\sum_m\{\phi^{\dagger}_m
\star\omega\star P_{-}\phi_m-
\varphi^{\dagger}_m\star\omega\star P_{+}\varphi_m\}
\end{displaymath}
and the current $\cJ_\mu^{(cov)}$ is defined by the identity 
\begin{displaymath}
\big(\cJ_\mu^{(cov)}\big)^{j}_{\;\,i}(x)=
-i\langle \big(\psi^i_{\beta}\star\bar{\psi}_{\alpha\,j}\big)(x)
\big(\gamma_{\mu}{\rm P}_{+}\big)_{\alpha\beta}\rangle.
\end{displaymath}
$\langle\cdots \rangle$ denotes the vacuum expectation value as given by 
\begin{displaymath}
\langle\cdots \rangle\,=\,
\frac{\int  d\bar{\psi}d\,\psi\quad\langle\cdots \rangle\quad e^{-{\rm S}[\A]}}
{\int  d\bar{\psi}d\,\psi\quad  e^{-{\rm S}[\A]}},
\end{displaymath}
with the fermionic measure $d\bar{\psi}d\,\psi$ as defined by 
eq.~(\ref{measure}).

As it stands, the right hand side of eq.~(\ref{formalanom}) is ill-defined; 
we shall obtain a well-defined object out of it by using 
the Gaussian cut-off furnished by the eigenvalues $\lambda^2_m$ in 
eq.~(\ref{definitions}). This well-defined object, which we shall denote
by $\cA[\omega,\A]^{(cov)}$, is the covariant form of the gauge anomaly:
\begin{displaymath}
\begin{array}{l}
{\cA[\omega,\A]^{(cov)}=\mbox{lim}_{\Lambda\rightarrow\infty}\;
\idxn\sum_m\;
e^{-\frac{\lambda_m^2}{\Lambda^2}}\Big\{\varphi_m^{\dagger}\star\omega\star
{\rm P}_{+}\varphi_m
-\phi_m^{\dagger}\star\omega\star
{\rm P}_{-}\phi_m\Big\}}\\[9pt]
{\phantom{\cA[\omega,\A]^{(cov)}}=
\mbox{lim}_{\Lambda\rightarrow\infty}\;
\idxn\sum_m\;\omega\star\Big\{
\big({\rm P}_{+}e^{-\frac{\lambda_m^2}{\Lambda^2}}
\varphi_m\big)\star\varphi_m^{\dagger}-
\big({\rm P}_{-}e^{-\frac{\lambda_m^2}{\Lambda^2}}
\phi_m\big)\star\phi_m^{\dagger}\Big\}}\\[9pt]
{\phantom{\cA[\omega,\A]^{(cov)}}=
\mbox{lim}_{\Lambda\rightarrow\infty}\;
\idxn\sum_m\;\omega\star\Big\{
\big({\rm P}_{+}e^{-\frac{(i\hat{D}(\A))^{\dagger}i\hat{D}(\A)}{\Lambda^2}}
\varphi_m\big)\star\varphi_m^{\dagger}-
\big({\rm P}_{-}e^{-\frac{i\hat{D}(\A)(i\hat{D}(\A))^{\dagger}}{\Lambda^2}}
\phi_m\big)\star\phi_m^{\dagger}\Big\}}\\[9pt]
{\phantom{\cA[\omega,\A]^{(cov)}}=
\mbox{lim}_{\Lambda\rightarrow\infty}\;
\idxn{\rm Tr}\;\omega\star\idpn\,{\rm tr}\;\Big\{
\Big(\gamma_{2n+1}\,e^{-\frac{(i\Dirac(\A))^2}{\Lambda^2}}\,e^{ipx}\Big)\star
e^{-ipx}\Big\}.}\\[9pt]
\end{array}
\end{displaymath}
The last line of the previous equation is obtained by changing to a plane-wave 
basis. In this last line $i\Dirac(\A)=i(\prslash+\Aslash\star)$ denotes 
the Dirac operator, and ${\rm Tr}$ and ${\rm tr}$ stand for traces over 
the $U(N)$ and Dirac matrices, respectively. 

Let ${\rm D(\A)}_{\mu}=\partial_{\mu}+\A_{\mu}\star$. Taking into account that 
$(i\Dirac(\A))^2=-({\rm D(\A)}^{\mu}{\rm D(\A)}_{\mu}+
\frac{1}{2}\gamma^{\mu}\gamma^{\nu}{\rm F}_{\mu\nu})$ and that  
${\rm D(\A)}^{\mu}{\rm D(\A)}_{\mu}(f\star e^{ipx})=
((ip_{\mu}+\partial_{\mu}+\A_{\mu}\star)^2\,f)\star e^{ipx}$, and that the
Moyal product is associative,  one can  show that
\begin{displaymath}
\begin{array}{l}
{\mbox{lim}_{\Lambda\rightarrow\infty}\;
\idxn {\rm Tr}\;\omega\star\idpn\,{\rm tr}\;\Big\{
\Big(\gamma_{2n+1}\,e^{-\frac{(i\Dirac(\A))^2}{\Lambda^2}}\,e^{ipx}\Big)\star
e^{-ipx}\Big\}=}\\[9pt]
{\idxn\, {\rm Tr}\;\omega\star\sum_{m=0}^{\infty}\Big\{ 
\mbox{lim}_{\Lambda\rightarrow\infty}\frac{1}{m!\Lambda^{2m}}\idpn\, 
e^{-\frac{p^2}{\Lambda^2}}}\\[9pt]
{\,{\rm tr}
\Big\{\gamma_{2n+1}\,\big[(\partial_{\mu}+\A_{\mu}\star)^2+2ip_{\mu}(\partial_{\mu}+\A_{\mu}\star)+\frac{1}{2}
\gamma^{\mu}\gamma^{\nu}{\rm F}_{\mu\nu}\star\big]^m\,\unit\Big\}\star e^{ipx}
\star e^{-ipx}\Big\}.}\\[9pt]
\end{array}
\end{displaymath}
Putting it all together, we conclude that 
\begin{displaymath}
\cA[\omega,\A]^{(cov)}\,=\,
\frac{i^n}{(4\pi)^n\,n!}\,\varepsilon^{\mu_1\cdots\mu_{2n}}\,{\rm Tr}\,\idxn\,
\omega\star\F_{\mu_1\mu_2}\star\cdots\star\F_{\mu_{2n-1}\mu_{2n}}(x).
\end{displaymath}
It is clear that $\cA[\omega,\A]^{(cov)}$ does not satisfy the Wess-Zumino 
consistency conditions
\begin{displaymath}
\delta_{\omega_{\,1}}\cA(\omega_{2},\A)\;-\;
\delta_{\omega_{\,2}}\cA(\omega_1,\A)\;=\;
\cA([\omega_1,\omega_2],\A).
\end{displaymath}
Hence, $\cJ^{(con)}_{\mu}(x)$ cannot be expressed as the derivative of 
effective action $\W[\A]\,=\,-\ln \Z[\A]$ with respect to the gauge field.

\section{The covariant form of the gauge anomaly: bi-fundamental and adjoint chiral matter}

We shall consider  a bi-fundamental~\cite{Terashima:2000xq} chiral fermion
$\psi^{i}_{R\,j}={\rm P}_{+}\psi^{i}_{\,j}$, $i=1,\cdots, N$ and $j=1,\cdots, M$  coupled  to a $U(N)$ gauge field, say,  $\A_{\mu}$, 
and a $U(M)$ gauge field, say, $\B_{\mu}$. The classical action of this
theory reads
\begin{equation}
{\rm S}\;=\;\idxn\; \bar{\psi}^{j}_{\;i}
\star(i\hat{D}(\A,\B)\psi)^{i}_{\;j}.
\label{biaction}
\end{equation}
The elliptic operator $i\hat{D}(\A,\B)$ acts on the bi-fundamental 
Dirac spinor $\psi^{i}_{\;j}$ as follows
\begin{displaymath}
(i\hat{D}(\A,\B)\psi)^{i_1}_{\; j_1}\,=\,i\big(
\prslash\,\delta^{i_1}_{\; i_2}\delta_{j_1}^{\; j_2}
+ \A^{i_1}_{\mu\,i_2}\star\,\delta_{j_1}^{\; j_2}\gamma^{\mu}{\rm P}_{+}
- \delta^{i_1}_{\; i_2} \star\B_{\mu\,j_1}^{j_2}\gamma^{\mu}{\rm P}_{+}\big)
\psi^{i_2}_{\; j_2}.
\end{displaymath}
We are using the following notation with regard to the $\star$-product: 
$\A_{\mu}\star\big)\psi\equiv \A_{\mu}\star\psi$ and 
$\star \B_{\mu}\big)\psi\equiv \psi \star \B_{\mu} $. Throughout this section, 
the $i$-indices run from $1$ to $N$ and the $j$-indices run from $1$ to $M$. 
$\A_{\mu}$ and $\B_{\mu}$ are anti-Hermitian matrices.

The action in eq.~(\ref{biaction}) is invariant under the following 
infinitesimal gauge  transformations
\begin{equation}
\begin{array}{l}
{(\delta_{(\omega,\chi)} \psi)^{i_1}_{\;j_1}=
\bigg(\omega^{i_1}_{\;i_2}\star {\rm P}_{+}\psi^{i_2}_{\;j_1}
-{\rm P}_{+}\psi^{i_1}_{\;j_2}\star\chi^{j_2}_{\;j_1}\bigg),}\\[9pt]
{(\delta_{(\omega,\chi)} \bar{\psi})^{j_1}_{\;i_1}=
-\bigg(\bar{\psi}^{j_1}_{\;i_2}\star\omega^{i_2}_{\;i_1}{\rm P}_{-} 
-\chi^{j_1}_{\;j_2}\star\bar{\psi}^{j_2}_{\;j_1}{\rm P}_{-}\bigg),}\\[9pt]
{\big(\delta_{\omega} \A_{\mu}\big)^{i_1}_{\;i_2} = 
-\partial_{\mu}\omega^{i_1}_{\;i_2}-\,\A^{i_1}_{\mu\,i_3}\star
\omega^{i_3}_{\;i_2}+
\,\omega^{i_1}_{\;i_3}\star\A^{i_3}_{\mu\,i_2},}\\[9pt]
{\big(\delta_{\chi} {\B}_{\mu}\big)^{j_1}_{\;j_2} =- 
\partial_{\mu}\chi^{j_1}_{\;j_2}-
\,{\B}^{j_1}_{\mu\,j_3}\star\chi^{j_3}_{\;j_2}+
\,\chi^{j_1}_{\;j_3}\star{\B}^{j_3}_{\mu\,j_2},}\\[9pt]
\end{array}
\label{bitrans}
\end{equation}
where 
$\omega^{i_1}_{\;i_2}=-\omega^{*\,i_2}_{\quad i_1}$, $i_1 , i_2=1,\cdots,N$, 
and $\chi^{j_1}_{\;j_2}=-\chi^{*\,j_2}_{\quad j_1}$, $j_1, j_2 =1,\cdots,M$, 
are the infinitesimal gauge transformation parameters.

Following the strategy developed in the previous section, we obtain
\begin{displaymath}
\Z[\A, \B] \;\equiv\;\int  d\bar{\psi}d\,\psi\quad e^{-{\rm S}[\A,\B]}\,=\,
\int\,\prod_{m}\,d{\bar b}_m da_m\; e^{-\sum_{n}\lambda_m {\bar b}_m a_m}\,=\,
\prod_{m}\lambda_m[\A,\B].
\end{displaymath}
The Grassmann variables, $a_m$ and ${\bar b}_m$, are given now by
the expansions $\psi=\sum_{m}a_m\varphi_m$ and $\bar{\psi}=\sum_{m}\bar{b}_m\phi_m^{\dagger}$; $\varphi_m$ and $\phi_m$ being the eigenvectors solving 
the eigenvalue problems and satisfying the identities that one gets by  replacing in eq.~(\ref{definitions}) $i\hat{D}(\A)$  with $i\hat{D}(\A,\B)$. 
$\lambda_m^2[\A,\B]$ are the eigenvalues of the problems so obtained. These 
eigenvalues are invariant under the infinitesimal gauge transformations of eq.~(\ref{bitrans}). Hence, the zeta 
regularization version of $\Z[\A, \B]$ above is gauge invariant.

Proceeding as in the previous section, one obtains the covariant form the 
gauge anomaly equation for bi-fundamental chiral matter:
\begin{equation}
\idxn\,\Big[\omega^{i_1}_{\;i_2}\star\, 
\big(D_{\mu}[\A]\,\cJ^{(\A,\,cov)}_\mu\big)^{i_2}_{\;\,i_1}\,+\,
\chi^{j_1}_{\;j_2}\star\, 
\big(D_{\mu}[\B]\,\cJ^{(\B,\, cov)}_\mu\big)^{j_2}_{\;\,j_1}
\,=\,\cA[\omega,\chi;\A,\B]^{(cov)}.
\label{bianom}
\end{equation}
Here the currents $\cJ^{(\A,\,cov)}_\mu$ and 
$\cJ^{(\B,\,cov)}_\mu$ 
are  defined, respectively, by the identities 
\begin{equation}
\begin{array}{c}
{\big(\cJ^{(\A,\,cov)}_\mu\big)^{i_1}_{\;\,i_2}(x)=
-i\langle \big(\psi^{i_1}_{\beta\,j_1}\star\bar{\psi}^{j_1}_{\alpha\,i_2}\big)(x)
\big(\gamma_{\mu}{\rm P}_{+}\big)_{\alpha\beta}\rangle,
}\\[9pt]
{\big(\cJ^{(\B,\,cov)}_\mu\big)^{j_1}_{\;\,j_2}(x)=
-i\langle \big(\bar{\psi}^{j_1}_{\alpha\,i_1}\star \psi^{i_1}_{\beta\,j_2}\big)(x)
\big(\gamma_{\mu}{\rm P}_{+}\big)_{\alpha\beta}\rangle;}\\[9pt]
\end{array}
\label{bicurrents}
\end{equation}
and $\cA[\omega,\chi;\A,\B]^{(cov)}$ is given by
\begin{equation}
\begin{array}{l}
{\cA[\omega,\chi;\A,\B]^{(cov)}=\mbox{lim}_{\Lambda\rightarrow\infty}\;
\idxn\sum_m\;
e^{-\frac{\lambda_m^2[\A,\B]}{\Lambda^2}}\Big\{\varphi_m^{\dagger}\star\omega\star
{\rm P}_{+}\varphi_m
-\phi_m^{\dagger}\star\omega\star
{\rm P}_{-}\phi_m\Big\}-}\\[9pt]
{\phantom{\cA[\omega,\chi;\A,\B]^{(cov)}=\;}
\mbox{lim}_{\Lambda\rightarrow\infty}\;
\idxn\sum_m\;
e^{-\frac{\lambda_m^2[\A,\B]}{\Lambda^2}}\Big\{\chi\star\varphi_m^{\dagger}
\star{\rm P}_{+}\varphi_m
-\chi\star\phi_m^{\dagger}\star{\rm P}_{-}\phi_m\Big\}.}\\[9pt]
\end{array}
\label{bianomaly}
\end{equation}

Let $i\Dirac(\A,\B)=i(1_{N\times N}\ot 1_{M\times M}\,
\prslash\,+\,1_{M\times M}\ot \Aslash\star\,+\,
\star\Bslash\,\ot 1_{N\times N}\;)$ the Dirac
operator acting on the bi-fundamental Dirac  spinor $\psi^{i}_{\; j}$, 
where $1_{N\times N}$ and $1_{M\times M}$ denote the unit matrices. It
can be shown that the right hand side of eq.~(\ref{bianomaly}) can be written
as follows
\begin{equation}
\begin{array}{l}
{\mbox{lim}_{\Lambda\rightarrow\infty}\;
\idxn\;{\rm Tr}_{\,\MM_{N\times N}}\,{\rm Tr}_{\,\MM_{M\times M}}\;\omega\star\idpn\,{\rm tr}\;\Big\{
\Big(\gamma_{2n+1}\,e^{-\frac{(i\Dirac(\A,\B))^2}{\Lambda^2}}\,e^{ipx}\Big)\star
e^{-ipx}\Big\}-}\\[9pt]
{\mbox{lim}_{\Lambda\rightarrow\infty}\;
\idxn\;{\rm Tr}_{\,\MM_{N\times N}}\,{\rm Tr}_{\,\MM_{M\times M}}\;
\chi\star\idpn\,{\rm tr}\;\Big\{
e^{-ipx}\star\Big(\gamma_{2n+1}\,e^{-\frac{(i\Dirac(\A,\B))^2}
{\Lambda^2}}\,e^{ipx}\Big)\Big\}.}\\[9pt]
\end{array}
\label{preparecomputation}
\end{equation}
${\rm tr}$ denotes the trace over the gamma matrices, and 
${\rm Tr}_{\,\MM_{N\times N}}$ and
${\rm Tr}_{\,\MM_{N\times N}}$ stand for the trace over  $N\times N$  and 
$M\times M$ complex matrices, respectively. 
It is not difficult to see that
\begin{displaymath}
\begin{array}{l}
{(\Dirac(\A,\B))^2=(1_{N\times N}\ot \,1_{M\times M}\,\partial_{\mu}\,+
\,1_{M\times M}\ot\,\A_{\mu}\star\,-\, \star\B_{\mu}\ot 1_{N\times N})^2\,+}\\[9pt]
{\phantom{(\Dirac(\A,\B))^2=}\frac{1}{2}\gamma^{\mu}\gamma^{\nu}\,(1_{M\times M}\ot F_{\mu\nu}[\A]\star\,-\,\star 
F_{\mu\nu}[\B]\ot 1_{N\times N}).}
\end{array}
\end{displaymath}
In the previous equation $\A_{\mu}\star$, $\star\B_{\mu}$,  
$F_{\mu\nu}[\A]\star$ and $\star F_{\mu\nu}[\B]$ are to be understood as 
operators acting on a appropriate matrix valued functions $f$ and $g$ as 
follows: $\A_{\mu}\star)f=\A_{\mu}\star f$, 
$\star\B_{\mu})g=g \star\B_{\mu}$, 
$F_{\mu\nu}[\A]\star)f=F_{\mu\nu}[\A]\star f$, 
$\star F_{\mu\nu}[\B])g=g \star F_{\mu\nu}[\B]$. $f$ takes values on  
$N\times N$ complex matrices and $g$ takes values on the $M\times M$ complex 
matrices. $F_{\mu\nu}[\A]$ and
$F_{\mu\nu}[\B]$ are the field strengths of $\A$ and $\B$, respectively. 
$1_{N\times N}$ and $1_{M\times M}$ are, respectively, the identity matrices 
of rank $N$ and $M$. Now, taking into  account that 
$
e^{ipx}\star f(x)\star e^{-ipx}\,=\,f(x+\theta p)$,
with $(\theta p)^{\mu}=\theta^{\mu\nu}p_{\nu}$, it is not difficult to show 
that eq.~(\ref{preparecomputation}) can be cast into the form
\begin{equation}
\begin{array}{l}
{\idxn\,{\rm Tr}_{\,\MM_{N\times N}}{\rm Tr}_{\,\MM_{M\times M}} \;
\omega\star\sum_{m=0}^{\infty}\Bigg\{ 
\mbox{lim}_{\Lambda\rightarrow\infty}\frac{1}{m!\Lambda^{2m}} 
\,\idpn\,e^{-\frac{p^2}{\Lambda^2}}}\\[9pt]
{{\rm tr}\Big\{\gamma_{2n+1}\,\Big[\big[1_{N\times N}\ot 1_{M\times M}\,
\partial_{\mu}+1_{M\times M}\ot A_{\mu}(x)\star -
\star\B_{\mu}(x+\theta p)\,\ot 1_{N\times N}\big]^2+}\\[9pt]
{\phantom{{\rm tr}\Big\{\gamma_{2n+1}\,\Big[\;}2ip_{\mu}
\big[1_{N\times N}\ot 1_{M\times M}\,
\partial_{\mu}+1_{M\times M}\ot\A_{\mu}(x)\star -
\star\B_{\mu}(x+\theta p)\,\ot 1_{N\times N}\big]+}\\[9pt]
{\phantom{\Big\{\gamma_{2n+1}\,\Big[[\;}\frac{1}{2}
\gamma^{\mu}\gamma^{\nu}\big[1_{M\times M}\ot {\rm F}_{\mu\nu}[\A](x)\star- \star{\rm F}_{\mu\nu}[\B](x+\theta p)\,\ot 1_{N\times N}
\big]\Big]^m\,\unit\Big\}\star e^{ipx}
\star e^{-ipx}\Bigg\}-}\\[9pt]
{\idxn\,{\rm Tr}_{\,\MM_{N\times N}}{\rm Tr}_{\,\MM_{M\times M}} \;
\chi\star\sum_{m=0}^{\infty}\Bigg\{
\mbox{lim}_{\Lambda\rightarrow\infty}\frac{1}{m!\Lambda^{2m}}  
\,\idpn\,e^{-\frac{p^2}{\Lambda^2}}}\\[9pt]
{{\rm tr}\Big\{\gamma_{2n+1}\, e^{-ipx}\star e^{ipx}\star
\Big[\big[1_{N\times N}\ot 1_{M\times M}\,\partial_{\mu}+ 1_{M\times M}\ot A_{\mu}(x-\theta p)\star -\star\B_{\mu}(x)\,\ot 1_{N\times N}\big]^2+}\\[9pt]
{\phantom{{\rm tr}\Big\{\gamma_{2n+1}\, e^{-ipx}\star e^{ipx}\star
\Big[\big[} 2ip_{\mu}\big[1_{N\times N}\ot 1_{M\times M}\,
\partial_{\mu}+ 1_{M\times M}\ot \A_{\mu}(x-\theta p)\star -
\star\B_{\mu}(x)\,\ot 1_{N\times N}\big]+}\\[9pt]
{\phantom{\Big\{\gamma_{2n+1}\,\Big[[\;}\frac{1}{2}
\gamma^{\mu}\gamma^{\nu}\big[1_{M\times M}\ot {\rm F}_{\mu\nu}[\A](x-\theta p)\star- \star{\rm F}_{\mu\nu}[\B](x)\,\ot 1_{N\times N}\big]\Big]^m\,\unit\Big\}\Bigg\}.}\\[9pt]
\end{array}
\end{equation}
$\unit$ is the symbol for the unit function on $\RR^{2n}$. Let us now recall that
${\rm tr}\,(\gamma_{2n+1}\gamma^{\mu_1}\cdots \gamma^{\mu_k})=0$, if 
$k<2n$; and that for a given value of $m$ the limit 
${\lim}_{\Lambda\rightarrow\infty}$ above vanishes, if the number of powers 
of $\Lambda$ turns negative upon  rescaling $p$ to $\Lambda p$. Keeping these 
two results in mind, one can show that 
the expression in the previous equation is equal to
\begin{equation}
\begin{array}{l}
{\idxn\,{\rm Tr}_{\,\MM_{N\times N}}{\rm Tr}_{\,\MM_{M\times M}} \;
\omega\star\Bigg\{
\mbox{lim}_{\Lambda\rightarrow\infty}\frac{i^n}{n!} 
\,\idpn\,e^{-p^2}\varepsilon^{\mu_1\cdots\mu_{2n}}}\\[9pt]
{\big[{\rm F}_{\mu_1\mu_2}[\A](x)\star - 
\star{\rm F}_{\mu_1\mu_2}[\B](x+\Lambda\,\theta p)\big]
\cdots\big[{\rm F}_{\mu_{2n-1}\mu_{2n}}[\A](x)\star - 
\star{\rm F}_{\mu_{2n-1}\mu_{2n}}[\B](x+\Lambda\,\theta p)\big]\,\unit\Bigg\}
-}\\[9pt]
{\idxn\,{\rm Tr}_{\,\MM_{N\times N}}{\rm Tr}_{\,\MM_{M\times M}} \;
\chi\star\Bigg\{ 
\mbox{lim}_{\Lambda\rightarrow\infty}\frac{i^n}{n!} 
\,\idpn\,e^{-p^2}\varepsilon^{\mu_1\cdots\mu_{2n}}}\\[9pt]
{\big[{\rm F}_{\mu_1\mu_2}[\A](x-\Lambda\,\theta p)\star - 
\star{\rm F}_{\mu_1\mu_2}[\B](x)\big]\cdots
\big[{\rm F}_{\mu_{2n-1}\mu_{2n}}[\A](x-\Lambda\,\theta p)\star - 
\star{\rm F}_{\mu_{2n-1}\mu_{2n}}[\B](x)\big]\,\unit\Bigg\}.}\\[9pt]
\end{array}
\label{bimixed}
\end{equation}
Generally speaking, in noncommutative quantum field theory, the 
limits $\Lambda\rightarrow\infty$ and $\theta p\rightarrow 0$ do not 
commute as a consequence of the intriguing 
UV/IR mixing~\cite{Minwalla:2000px}. 
Then, to define the renormalized theory, one  has make a choice regarding 
the order of these limits. One would like to obtain the renormalized 
noncommutative theory at $\theta p = 0$ by taking the limit $\theta p\rightarrow 0$ of renormalized one at $\theta p \neq 0$. Hence, we shall take  
the limit $\Lambda\rightarrow\infty$ first and then take 
the limit $\theta p\rightarrow 0$. Now, the gauge fields $\A_{\mu}$ and 
$\B_{\mu}$ satisfy the boundary conditions $\F_{\mu\nu}[\A](y)\rightarrow 0$ 
and $\F_{\mu\nu}[\B](y)\rightarrow 0$ as $|y|\rightarrow \infty$. It is thus 
plain  that ~(\ref{bimixed}) is equal to
\begin{equation}
\begin{array}{l}
{M\,\frac{i^n}{(4\pi)^n\,n!}\,\varepsilon^{\mu_1\cdots\mu_{2n}}\,
{\rm Tr}_{\,\MM_{N\times N}}\,\idxn\,\omega\star
\F_{\mu_1\mu_2}[\A]\star\cdots\star\F_{\mu_{2n-1}\mu_{2n}}[\A](x)\,-}\\[9pt]
{N\,\frac{(-i)^n}{(4\pi)^n\,n!}\,\varepsilon^{\mu_1\cdots\mu_{2n}}\,{\rm Tr}_{\,\MM_{M\times M}}\,\idxn\,\omega\star
\F_{\mu_1\mu_2}[\B]\star\cdots\star\F_{\mu_{2n-1}\mu_{2n}}[\B](x).}\\[9pt]
\end{array}
\label{bicompu}
\end{equation}
Putting it all together (see~(\ref{bianom})--(\ref{bicompu})), we conclude that the covariant form of the anomaly
$\cA[\omega,\chi;\A,\B]^{(cov)}$ reads thus
\begin{equation}
\begin{array}{l}
{\cA[\omega,\chi;\A,\B]^{(cov)}\,=\, M\,\frac{i^n}{(4\pi)^n\,n!}\,\varepsilon^{\mu_1\cdots\mu_{2n}}\,
{\rm Tr}_{\,\MM_{N\times N}}\,\idxn\,\omega\star
\F_{\mu_1\mu_2}[\A]\star\cdots\star\F_{\mu_{2n-1}\mu_{2n}}[\A](x)\,-}\\[9pt]
{\phantom{\cA[\omega,\chi;\A,\B]^{(cov)}\,=\,\;}N\,\frac{(-i)^n}{(4\pi)^n\,n!}\,\varepsilon^{\mu_1\cdots\mu_{2n}}\,{\rm Tr}_{\,\MM_{M\times M}}\,\idxn\,\omega\star
\F_{\mu_1\mu_2}[\B]\star\cdots\star\F_{\mu_{2n-1}\mu_{2n}}[\B](x).}\\[9pt]
\end{array}
\label{bicovanom}
\end{equation}
Notice that as in the consistent 
case~\cite{Martin:2000qf, Intriligator:2001yu} there are no
mixed anomalies. Also notice that if in~ (\ref{bimixed}) we set $\theta = 0$ 
before sending $\Lambda$ to $\infty$, ie, we go to commutative space,  
the mixed anomalies pop-up back; and that it is the characteristic  
half-dipole structure of the charged degrees of freedom of the 
noncommutative field theories 
--see the ${\rm F}_{\mu_{i}\mu_{i+1}}[\A](x-\Lambda\,\theta p)$ and 
${\rm F}_{\mu_{i}\mu_{i+1}}[\B](x+\Lambda\,\theta p)$ terms in the mixed contributions of~ (\ref{bimixed})-- which is responsible for 
the lack of mixed anomalies in the noncommutative arena. That charged degrees 
of freedom have a  half-dipole structure rather than a dipole structure
~\cite{Sheikh-Jabbari:1999vm, Bigatti:2000iz}  was 
unveiled in ref.~\cite{Alvarez-Gaume:2001bv}.

  The consistent form of the gauge anomaly for an adjoint right-handed fermion can be obtained by setting $\A\,=\,\B$ in eq.~(\ref{bicovanom}). We thus 
conclude that if $D=4m$ ($D$ is the space-time dimension) there is no gauge anomaly, but if $D=4m+2$, the anomaly is $2N$ times the anomaly in the 
fundamental representation.

\section{Redefinition  of currents}

In this section we shall show that there exists a $*$-polynomial, $\cX^{\mu}$, 
of $\A$ and $\F$, such that 
\begin{equation}
\cJ^{(con)}_{\mu}(x)=\cJ^{(cov)}_{\mu}(x)\,+\,\cX_{\mu}(x).
\label{redefinition}
\end{equation}
Here $\cJ^{(con)}_{\mu}(x)$ denotes the consistent gauge current for a 
fundamental right handed fermion  
--see ref.~\cite{Gracia-Bondia:2000pz, Bonora:2000he, Grisaru:2001sk}-- and
$\cJ^{(cov)}_{\mu}(x)$ stands for the corresponding covariant gauge current. 
In view of the results presented in the previous section,  
the generalization of the analysis we are about to begin to  bi-fundamental and/or adjoint right handed fermions is trivial. 

  To compute $\cX^{\mu}$ we shall adapt to the 
case at hand the techniques of ref.~\cite{Alvarez-Gaume:1985dr}. To do so we 
shall employ the  formalism of differential forms and BRST cohomology 
introduced  in ref.~\cite{Bonora:2000he}. 

Let $\cJ^{(con)}$ and  $\cJ^{(cov)}$   be the dual 
currents         
\begin{equation}
\begin{array}{l} 
{\cJ^{(con)}=\frac{1}{(2n-1)!}\,
\varepsilon^{\mu_1}_{\phantom{\mu_1}\mu_2\cdots\mu_{2n}}\,
\cJ_{\mu_1}^{(con)}\,dx^{\mu_2}\cdots dx^{\mu_{2n}},}\\[9pt]
{\cJ^{(cov)}=\frac{1}{(2n-1)!}\,
\varepsilon^{\mu_1}_{\phantom{\mu_1}\mu_2\cdots\mu_{2n}}\,
\cJ_{\mu_1}^{(cov)}\,dx^{\mu_2}\cdots dx^{\mu_{2n}}.}\\[9pt]
\end{array}
\label{dualcurrents}
\end{equation}
These currents are $(2n-1)$-differential forms in the sense of  
ref.~\cite{Bonora:2000he}. Let
$\C$ be the ghost zero-form introduced through the BRST transformations:
$sA=D\C=d\C+[\A,\C]$, $sc=\C\star\C$. $s$ is the BRST operator,  
$d$ is the exterior derivative and $\A=\A_{\mu}\,dx^{\mu}$. $s$ and $d$ 
satisfy $s^2\,=\,d^2\,=\,sd+ds\,=\,0$.
We  introduce next the two-form field-strength    
$\F=\frac{1}{2}\F_{\mu\nu}\,dx^{\mu}dx^{\nu}\,=\,d\A\,+\,\A\star\A$.
Then, $\cJ^{(con)}$ and $\cJ^{(cov)}$ are defined so that they 
satisfy, respectively, the consistent form and the covariant form
of gauge anomaly equation:
\begin{equation}
\int {\rm Tr}\ D\C\star\cJ^{(con)}\,=\,\cA(\C,\A)^{(con)}\qquad
\mbox{\rm  and}\qquad
\int {\rm Tr}\ D\C\star\cJ^{(cov)}\,=\,\cA(\C,\A)^{(cov)},
\label{anomalyindifforms}
\end{equation}
where
\begin{equation}
\cA(\C,\A)^{(con)}\,=\,\frac{i^n}{(2\pi)^n\,(n+1)!}\,\int\, \cQ^{1}_{\;\,2n}
\,(\C,\A,\F)
\label{conform}
\end{equation}
and 
\begin{equation}
\cA(\C,\A)^{(cov)}\,=\,\frac{i^n}{(2\pi)^n\,n!}\,\int\,
\bracl{\rm Tr}\;\C\star \F^n\bracr.
\label{covform}
\end{equation}
$\cQ^{1}_{2n}\,(\C,\A,\F)$, which can be obtained by solving the descent 
equations, reads  
\begin{equation}
\cQ^{1}_{\;\,2n}\,(\C,\A,\F)\,=\,
(n+1)\int_{0}^{1}dt\,(1-t)\,\sum_{k=0}^{n-1}\,\bracl{\rm Tr}\;
\C\star d(\F_t^{k}\star\A\star \F_t^{n-1-k})\bracr,
\label{Qform}
\end{equation}
with $\F_t=d\A_t+\A_t^2$ and $\A_t=t\A$. 
$\F^k$ denotes the $k$-th power of $\F$ with respect to the Moyal product. 
An expression like $\bracl
{\rm Tr}\;( \E_1 \star\E_2\star\cdots\star\E_m)\bracr$ denotes
the equivalence class obtained by imposing on the space of objects of the type 
${\rm Tr}\,( \E_1 \star\E_2\star\cdots\star\E_m)$ the relationship 
${\rm Tr}\;( \E_1 \star\E_2\star\cdots\star\E_m)\equiv 
(-1)^{k_m(k_1+\cdots+k_{m-1})}
{\rm Tr}\;( \E_m \star\E_1\star\cdots\star\E_{m-1})$. 
$\E_i$ denotes a form of degree  $k_i$. See ref.~\cite{Bonora:2000he} for 
further details. 

 To find $\cX$, 
\begin{equation}
{\cX}=\frac{1}{(2n-1)!}\,
\varepsilon^{\mu_1}_{\phantom{\mu_1}\mu_2\cdots\mu_{2n}}\,
\cX_{\mu_1}\,dx^{\mu_2}\cdots dx^{\mu_{2n}},
\label{Xform}
\end{equation}
such that $\cX_{\mu}$ satisfies eq.~(\ref{redefinition}), we shall first
show that $\cQ^{1}_{\;\,2n}\,(\C,\A,\F)$ in eq.~(\ref{Qform}) is also
given by the following equation 
\begin{equation}
\begin{array}{l}
{\cQ^{1}_{\;\,2n}\,(\C,\A,\F)\,=\,(n+1)\bracl{\rm Tr}\;\C\star \F^n\bracr\,-\,}
\\[9pt]
{\phantom{\cQ^{1}_{\;\,2n}\,(\C,\A,\F)\,=\,}
(n+1)\int_{0}^{1}dt\,\bracl{\rm Tr}\;\C\star D(
\sum_{k=0}^{n-1}\,\F_t^{k}\star\A_t\star \F_t^{n-1-k})\bracr.}\\[9pt]
\end{array}
\label{newexpression}
\end{equation}
It can be shown that the right hand side of eq.~(\ref{Qform}) is equal to
\begin{equation}
(n+1)\int_{0}^{1}dt\,\bracl{\rm Tr}\;\C\star \F_t^n\bracr\,+\,
(n+1)\int_{0}^{1}dt\,\sum_{k=0}^{n-1}\,(t-1)\,\bracl{\rm Tr}\;
(\F_t^{k}\star\A_t\star \F_t^{n-1-k}\star[\A,\C])\bracr.
\label{eqone}
\end{equation}
Now, taking into account that $(D_t-D)\C\,=\,(t-1)[\A,\C]$ and that 
$\bracl{\rm Tr} D\cO\bracr\,=\,d\bracl{\rm Tr} \cO\bracr$ for a form,  
$\cO$, of even degree; one readily shows that eq.~(\ref{eqone}) can
be turned into the following expression
\begin{displaymath}
\begin{array}{l}
{(n+1)\Big(\int_{0}^{1}dt\,\bracl{\rm Tr}\;\C\star \F_t^n\bracr\,
\,-\, d\,\int_{0}^{1}dt\,\sum_{k=0}^{n-1}\,\bracl{\rm Tr}\;
(\F_t^{k}\star\A_t\star \F_t^{n-1-k}\star\C)\bracr\,+\,}\\[9pt] 
{\int_{0}^{1}dt\,\sum_{k=0}^{n-1}\,\bracl{\rm Tr}\;
(\F_t^{k}\star D_t\A_t\star \F_t^{n-1-k}\star\C)\bracr\,-\,
\int_{0}^{1}dt\,\sum_{k=0}^{n-1}\,\bracl{\rm Tr}\;
(\F_t^{k}\star \A_t\star \F_t^{n-1-k}\star D\C)\bracr\Big).}\\[9pt] 
\end{array}
\end{displaymath}
Upon employing that $D_t\A_t=t\partial_t\F_t$ and that $\partial_t\F_t^n\,=\,
\sum_{k=0}^{n-1}\F_t^{k}\star\partial_t\F_t \star \F_t^{n-1-k}$, 
the previous equation can be converted into the following one
\begin{displaymath}
\begin{array}{l}
{(n+1)\Big(\int_{0}^{1}dt\,\bracl{\rm Tr}\;\C\star \F_t^n\bracr\,
\,-\, d\,\int_{0}^{1}dt\,\sum_{k=0}^{n-1}\,\bracl{\rm Tr}\;
(\F_t^{k}\star\A_t\star \F_t^{n-1-k}\star\C)\bracr\,+\,}\\[9pt] 
{\phantom{(n+1)\Big(}\int_{0}^{1}dt\,t\; \partial_t\;\bracl{\rm Tr}\;
(\F_t^n\star\C)\bracr\,-\,
\int_{0}^{1}dt\,\sum_{k=0}^{n-1}\,\bracl{\rm Tr}\;
(\F_t^{k}\star \A_t\star \F_t^{n-1-k}\star D\C)\bracr\Big).}\\[9pt] 
\end{array}
\end{displaymath}
Partial integration yields then 
\begin{displaymath}
\begin{array}{l}
{(n+1)\Big(\bracl{\rm Tr}\;\C\star \F^n\bracr\,-\,
\int_{0}^{1}dt\,\sum_{k=0}^{n-1}\,\bracl{\rm Tr}\;
(\F_t^{k}\star \A_t\star \F_t^{n-1-k}\star D\C)\bracr\,-\,}\\[9pt]
{\phantom{(n+1)\Big(} d\,\int_{0}^{1}dt\,\sum_{k=0}^{n-1}\,\bracl{\rm Tr}\;
(\F_t^{k}\star\A_t\star \F_t^{n-1-k}\star\C)\bracr\Big).}\\[9pt] 
\end{array}
\end{displaymath}
This equation and 
\begin{displaymath}
d\,\bracl{\rm Tr}\;
(\F_t^{k}\star\A_t\star \F_t^{n-1-k}\star\C)\bracr\,=\,
\bracl{\rm Tr}\;(D(\F_t^{k}\star\A_t\star \F_t^{n-1-k})\star\C)\bracr
-\bracl{\rm Tr}\;
(\F_t^{k}\star\A_t\star \F_t^{n-1-k}\star D\C)\bracr
\end{displaymath}
finally lead to eq.~(\ref{newexpression}).
 
We are now ready to compute $\cX$ so that eqs.~(\ref{redefinition}) and~ 
(\ref{Xform}) hold:
\begin{equation}
\begin{array}{l}
{\int {\rm Tr}\ \C\star D\cX\,=\,\int {\rm Tr}\ \C\star D\cJ^{(con)}\,-\,
\int {\rm Tr}\ \C\star D\cJ^{(cov)}}\\[9pt]
{\phantom{\int {\rm Tr}\ D\C\star\cX}
 \,=\,\frac{i^n}{(2\pi)^n\,(n+1)!}\,\int\,\Big( \cQ^{1}_{\;\,2n}
\,(\C,\A,\F)\,-(n+1)\,
\bracl{\rm Tr}\;\C\star \F^n\bracr\Big)}\\[9pt]
{\phantom{\int {\rm Tr}\ D\C\star\cX}
 \,=\,\frac{i^{(n+2)}}{(2\pi)^n\,n!}\,\int\,
\int_{0}^{1}dt\,\bracl{\rm Tr}\;\C\star D(
\sum_{k=0}^{n-1}\,\F_t^{k}\star\A_t\star \F_t^{n-1-k})\bracr.}\\[9pt]
\end{array}
\label{Xequation}
\end{equation}
To obtain the previous array of identities, eqs.~(\ref{anomalyindifforms})--
(\ref{newexpression}) are to be taken into account. In view of 
eq.~(\ref{Xequation}), we conclude that the following choice of $\cX$, 
\begin{equation} 
\cX\,=\,\frac{i^{(n+2)}}{(2\pi)^n\,n!}\,
\int_{0}^{1}dt\;
\sum_{k=0}^{n-1}\,\F_t^{k}\star\A_t\star \F_t^{n-1-k},
\label{ValueofX}
\end{equation}
would do the job. Notice that the result we have obtained is the naive 
$\star$-deformation of the ordinary expression without symmetrization.

\section{Currents and gauge transformations}

In this section we shall study the behaviour under gauge transformations 
of the consistent and covariant dual currents --$\cJ^{(con)}$ and
$\cJ^{(cov)}$ in eq.~(\ref{dualcurrents}), respectively. We shall employ the
techniques of ref.~\cite{Alvarez-Gaume:1985dr} and show that, when
there is an anomaly, the following equation does not hold
\begin{displaymath}
s\,\cJ^{(con)}\,=\,[\C,\cJ^{(con)}],
\end{displaymath} 
but the following equation does
\begin{equation}
s\,\cJ^{(cov)}\,=\,[\C,\cJ^{(cov)}].
\label{covlaw}
\end{equation}

 The consistent current is obtained from the effective action $\W[\A]$ by
functional differentiation of the latter, ie,
\begin{displaymath}
\delta\,\W[\A]\,=\,\int\,{\rm Tr}\,\delta\A\, \star\,\cJ^{(con)}.
\end{displaymath}
The operator $\delta$ is given by $\int\,\delta\A\star\frac{\delta}{\delta\A}$.
The BRST variation of infinitesimal one-form $\delta\A$ is defined to be
$s\,\delta\A\,=\,[\delta\A,\C]$. It can be readily seen that 
$s\delta=\delta s$.
We next introduce the anti-derivation $l$ as follows
$l\A\,=\,0$, $l\F\,=\,\delta\A.$ It can be shown that 
$l\,d\,+\,d\,l\,=\,\delta$, on the space of polynomials of $\A$ and $\F$ 
with respect to the Moyal product. 

The consistent form of the gauge anomaly equation runs thus in terms 
of $\W[\A]$:
\begin{displaymath}
s\W[\A]\,=\,\frac{i^n}{(2\pi)^n\,(n+1)!}\,\int\, \cQ^{1}_{\;\,2n}
\,(\C,\A,\F).  
\end{displaymath}
$\cQ^{1}_{\;\,2n}\,(\C,\A,\F)$ is given in eq.~(\ref{Qform}).  
Acting with $\delta$ on both sides of the previous equation, one obtains
\begin{equation}
\begin{array}{l}
{\frac{i^n}{(2\pi)^n\,(n+1)!}\,\int\, \delta\,\cQ^{1}_{\;\,2n}
\,(\C,\A,\F)\,=\,\delta\,s\W[\A]\,=\,s\,\delta\W[\A]
\,=\,s\,\int\,{\rm Tr}\,\delta\A\, \star\,\cJ^{(con)}}\\[9pt]
{\phantom{\frac{i^n}{(2\pi)^n\,(n+1)!}\,\int\, \delta\,\cQ^{1}_{\;\,2n}}
\,=\,s\,\int\,{\rm Tr}\,\delta\A\, \star\,\cJ^{(con)}
   \,=\,-\int\,{\rm Tr}\,\delta\A\star\big\{s\cJ^{(con)}\,-\,
[\C,\cJ^{(con)}]
\big\}.}\\[9pt]
\end{array}
\label{lackofcovar}
\end{equation}
Hence, the existence of the gauge anomaly 
--$\cQ^{1}_{\;\,2n}\,(\C,\A,\F)$ does not vanish-- prevents $\cJ^{(con)}$ 
from transforming covariantly. Notice that eq.~(\ref{lackofcovar}) tell us 
that the behaviour of $\cJ^{(con)}$ under gauge transformations is given  
by the anomaly. 

Let us finally show that the covariant current, $\cJ^{(cov)}$, obtained 
by subtracting $\cX$ in eq.~(\ref{ValueofX}) from $\cJ^{(con)}$, deserves that 
name indeed, ie, it satisfies eq.~(\ref{covlaw}). In view of 
eq.~(\ref{lackofcovar}), we just have to show that
\begin{displaymath}
\int\,{\rm Tr}\,\delta\A\star\big\{-s\cX\,+\,
[\C,\cX]\big\}\,=\,
\frac{i^n}{(2\pi)^n\,(n+1)!}\,\int\, \delta\,\cQ^{1}_{\;\,2n},
\,(\C,\A,\F).
\end{displaymath}
with $\cX$ given in eq.~(\ref{Xform}). Taking into account the (graded) 
cyclicity of $\int\,{\rm Tr}\, E_{k_1}\star\cdots E_{k_n}$, and 
that $t\delta\A=l\,\F_t$ and that $l\,\F_t^n = \sum_{k=0}^{n}\,
\F_t^{n-1-k}\star l\F_t\star\F_t^k$, with $\F_t\,=\,d\A_t\,+\,\A_t^2$; 
one shows that  
\begin{displaymath}
\begin{array}{l}
{\int\,{\rm Tr}\,\delta\A\star\big\{-s\cX\,+\,
[\C,\cX]\big\}\,=\,s\int\,{\rm Tr}\,\delta\A\star\cX}\\[9pt]
{\phantom{-\int\,{\rm Tr}\,\delta\A\star\big\{s\cX\,-\,
[\C,\cX]\big\}} -\frac{i^n}{(2\pi)^n\,(n+1)!}\,\int\,s\,l\,\big\{(n+1)\,
\int_{0}^{1}\,dt\;{\rm Tr} \A\star\F_t^n\;\big\}.}\\[9pt]
\end{array}
\end{displaymath}
It is plain that 
\begin{displaymath}
\int\,s\,l\,\big\{(n+1)\,
\int_{0}^{1}\,dt\;{\rm Tr} \A\star\F_t^n\;\big\}\,=\,
\int\,s\,l\,\big\{(n+1)\,
\int_{0}^{1}\,dt\;\bracl{\rm Tr} \A\star\F_t^n\bracr\;\big\}.
\end{displaymath}
Now, the descent equation formalism of ref.~\cite{Bonora:2000he} leads to  
$s\cQ^{0}_{\;\,2n+1}\,(\A,\F)\,=\,d\cQ^{1}_{\;\,2n}\,(\C,\A,\F)$
where $\cQ^{0}_{\;\,2n+1}\,(\A,\F)\,=\,(n+1)\,
\int_{0}^{1}\,dt\;\bracl{\rm Tr} \A\star\F_t^n\bracr$ and $\cQ^{1}_{\;\,2n}\,(\C,\A,\F)$ is given in eq.~(\ref{Qform}). Putting it all together, we get 
\begin{displaymath}
\begin{array}{l}
{\int\,{\rm Tr}\,\delta\A\star\big\{-s\cX\,+\,
[\C,\cX]\big\}\,=\,
-\frac{i^n}{(2\pi)^n\,(n+1)!}\,\int\,s\,l\,\cQ^{0}_{\;\,2n+1}\,(\A,\F)
\,=\,}\\[9pt]
{\frac{i^n}{(2\pi)^n\,(n+1)!}\,\int\,l\,s\,\cQ^{0}_{\;\,2n+1}\,(\A,\F)
\,=\,\frac{i^n}{(2\pi)^n\,(n+1)!}\,\int\,l\,
d\,\cQ^{1}_{\;\,2n}\,(\C,\A,\F)\,=\,}\\[9pt]
{\frac{i^n}{(2\pi)^n\,(n+1)!}\,\int\,(-dl\,+\,\delta)
\,\cQ^{1}_{\;\,2n}\,(\C,\A,\F)
\,=\,
\frac{i^n}{(2\pi)^n\,(n+1)!}\,\int\,\delta
\,\cQ^{1}_{\;\,2n}\,(\C,\A,\F).}\\[9pt]
\end{array}
\end{displaymath}

\section{Conclusions}

In this paper we have considered the noncommutative gauge anomaly for $U(N)$ 
gauge groups.
We have shown that the covariant form of gauge anomaly on noncommutative
$\RR^{2n}$ can be understood as the lack of invariance of the 
fermionic measure under chiral gauge transformations of the fermion fields. 
This lack of invariance is given by a non-trivial Jacobian, which when
defined by using an appropriate regularization  yields the covariant form 
of the anomaly. By using these path integral techniques, we have finally 
computed the convariant form of the gauge anomaly on $\RR^{2n}$ to show 
that it is given by a $\star$-polynomial of the gauge field strength.
The covariant form of the gauge anomaly on even dimensional space is thus 
seen to be given by an appropriate $\star$-deformation of the ordinary 
expression.   

We have  proved that one can trade the covariant form of the gauge
anomaly for the consistent one by adding to the covariant current a
$\star$-polynomial of the gauge field and the gauge field strength.
We have computed this polynomial explicitly.
Gauge anomalies are thus given by local expressions in the sense of 
noncommutative geometry --of course, these expressions are non-local from 
the ordinary quantum field theory point of view. 

We have seen that the gauge transformation properties of the consistent 
current are given by the consistent form of the gauge the anomaly. The 
existence of the gauge anomaly  prevents 
the consistent current from transforming covariantly, but allows  
the covariant current to  transform covariantly under 
gauge transformations of the gauge field.

It is worth stressing that in the course of our path integral computations 
 we have proved that on noncommutative $\RR^{2n}$ the covariant form 
of the gauge anomaly for bi-fundamental  
chiral matter  carries no mixed anomaly. For the noncommutative 
theory, the absence of these mixed anomalies is interpreted 
--see ref.~\cite{Intriligator:2001yu}  for an 
interpretation in terms of the Green-Schwarz mechanism-- as a consequence 
of the half-dipole structure~\cite{Alvarez-Gaume:2001bv} which is 
characteristic of the noncommutative charged fields. 
From the covariant form of the gauge anomaly for bi-fundamental chiral matter,
one readily obtains the covariant form of the anomaly form adjoint chiral 
fermions. If $D$ denotes the espace dimension, our results -for adjoint chiral right-handed fermion in the continuum-- run thus: there is no anomaly at $D=4m$; at $D=4m+2$, the anomaly is $2N$ times the fundamental anomaly. Interestingly enough one can formulate such theories on the lattice in an anomaly 
free manner for any even integer $D$~\cite{Nishimura:2001dq}. 

It is an interesting task to try to extend the results presented here to 
groups  other than the $U(N)$ groups. New techniques and ideas such us the 
ones introduced in refs.~\cite{Bonora:2000td, Jurco:2001rq} will be 
unavoidably needed, if one is to succeed. 

Finally, as we were writing the closing sentences in this paper, we 
became aware of ref.~\cite{Bonora:2001fa}. The results discussed above are
in complete harmony with the results presented in this last reference.

\end{document}